# Toward a Healthier Social Media Experience: Designing 'Inspiration' and 'Reality' Modes to Enhance Digital Well-Being for Generation Z

Sora Kang[1]

[1]Human-Computer Interaction + Design Lab, Seoul National University, Seoul, Korea

E-mail: sorakang@snu.ac.kr

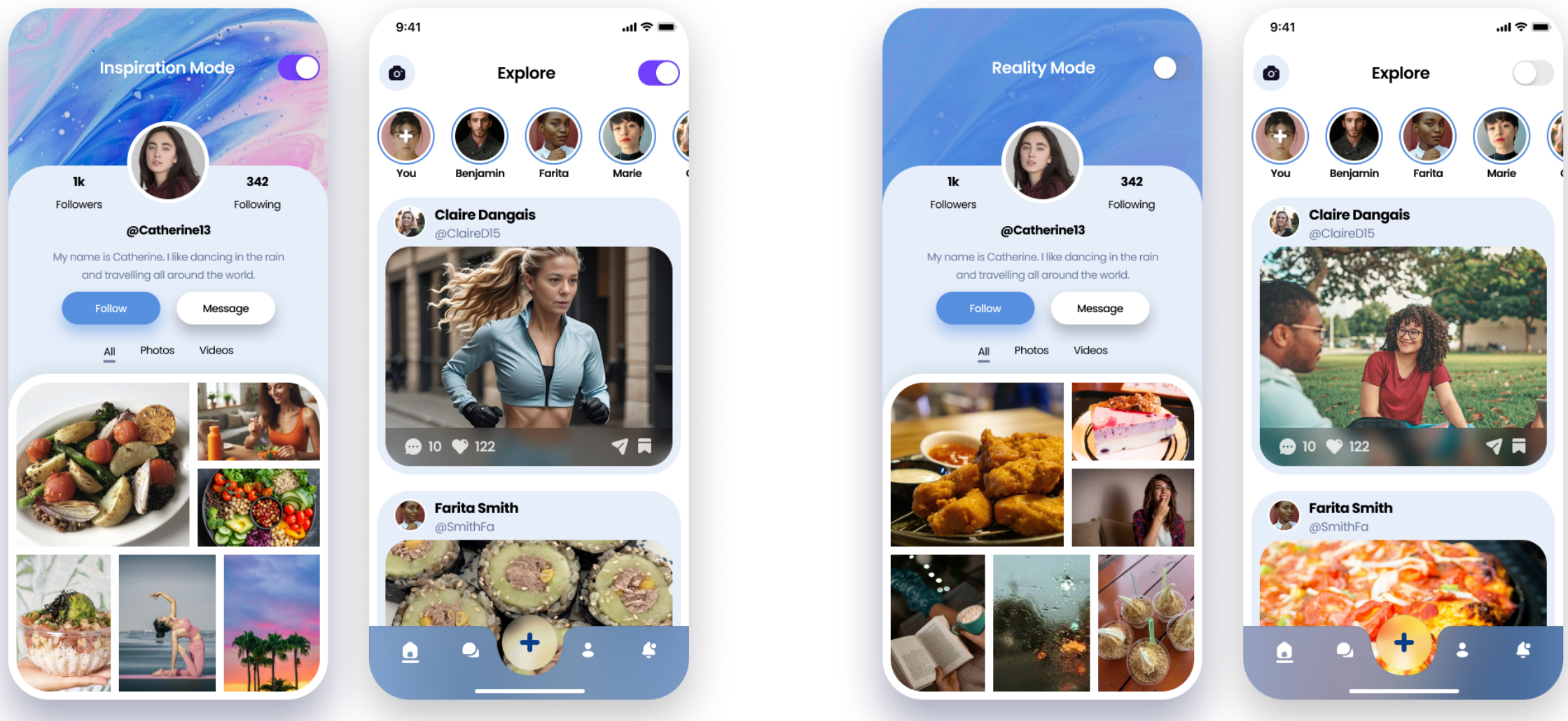

Fig. 1. Envisioned ‘Inspiration Mode’ and ‘Reality Mode’ Interface

*Abstract*—**This study presents a dual-mode interface design concept for social media platforms aimed at reducing social comparison in health-related content among Korean MZ (Millennials and Gen-Z) users. The proposed "Inspiration" and "Reality" modes allow users to toggle between curated, idealized posts and more realistic, candid content. This approach aims to alleviate negative psychological effects, such as decreased self-esteem and body dissatisfaction. The pre-study outlines the design framework and discusses potential implications for user satisfaction, perceived authenticity, and mental well-being.**

*Keywords*— **User Interface Design, Healthy Pleasure, MZ Generation, Social Media, Social Comparison**

## I. INTRODUCTION

The rise of **social media** has significantly transformed how people share and consume health-related content, especially among the younger generations[1]. In South Korea, the MZ generation (Millennials and Gen-Z), who are known for their digital literacy and active social media engagement, have embraced the trend of **“Healthy Pleasure”**[2], [3]. This trend prioritizes wellness practices that bring joy and sustainability over rigorous, discipline-focused routines. Popular practices include sharing workout completion posts using hashtags like #오운완 (*o-un-wan,* 오늘 운동 완료/*Workout done today*), documenting low-calorie meal choices, and showcasing personal fitness transformations. These activities not only foster a sense of community but also serve as a source of motivation and inspiration for users seeking to improve their health and well-being.

Despite the positive aspects of engaging with health content online, there is a growing concern about the negative psychological effects associated with social comparison on social media platforms[4], [5]. According to Festinger’s Social Comparison Theory(1954), individuals have an innate tendency to evaluate their own abilities and self-worth by comparing themselves to others[6]. On social media, where content is often curated and idealized, users are more likely to engage in upward comparisons, measuring themselves against posts that highlight the most polished and successful aspects of others’ health journeys. This can lead to feelings of inadequacy, reduced self-esteem, and increased body dissatisfaction, particularly among younger users who are still developing their self-identity.

Research has shown that exposure to idealized health and fitness imagery on platforms like Instagram and TikTok can exacerbate the pressures of social comparison, contributing to negative mental health outcomes. Studies by Turner (2015) indicate that Gen-Z users, in particular, are vulnerable to the effects of social comparison due to their preference for authenticity but simultaneous attraction to aspirational, visually appealing content[7]. This paradox creates a challenging environment for users who seek inspiration while also desiring relatable, genuine content.

To address these issues, recent work in the field of **Human-Computer Interaction (HCI)** has explored interface design strategies that prioritize user well-being. Techniques such as **self-reflection prompts**, **customizable content filters**, and **mood-based content adjustments** have shown promise in mitigating the negative impacts of social comparison by offering users greater control over their content consumption[8], [9], [10]. Building on these findings, the current study proposes a **dual-mode interface design** featuring **"Inspiration"** and **"Reality"** modes as a novel intervention to balance motivation with relatability in health-related content.

This study aims to evaluate the impact of this dual-mode interface on the **user experience**, particularly focusing on perceptions of authenticity, changes in self-esteem, and tendencies towards social comparison. It also considers the preferences and behaviors of the Korean MZ generation, a group that is heavily engaged in social media and increasingly interested in health and wellness trends. The findings from this pre-study could offer valuable insights into the design of more user-centric and well-being-oriented social media platforms, contributing to the broader discourse on digital health and mental wellness.

In summary, the proposed research seeks to answer the following key questions:

1. How does the dual-mode interface affect users' perceptions of health-related content in terms of authenticity and relatability?
2. What impact does the use of "Inspiration" and "Reality" modes have on users' self-esteem and social comparison tendencies?
3. How do Korean MZ users respond to this design intervention, and what are their preferences and behaviors when interacting with the dual-mode interface?

By exploring these questions, this study aims to contribute to the ongoing efforts in HCI to design digital platforms that support healthier and more balanced online experiences, aligning with the principles of **Healthy Pleasure** and promoting mental well-being among social media users.

## II. METHODS

This pre-study involves developing a social media prototype that features a dual-mode interaction design ("Inspiration" and "Reality" modes). The aim is to explore the effects of this design on reducing negative social comparison among Korean MZ users. This section provides a detailed outline of the prototype design, user testing procedure.

### A. Prototype Design

The prototype includes two distinct viewing modes to accommodate different user needs and preferences:

- **Inspiration Mode**: This mode features curated, idealized health and fitness content, including transformation photos, polished workout images, and success stories. It is designed to provide users with aspirational and motivational content, aligning with the typical presentation seen on popular fitness social media platforms.
- **Reality Mode**: This mode presents realistic, unfiltered posts that highlight the challenges and candid moments of health journeys. It includes everyday workout struggles, failures, and more authentic user-generated content. The aim is to reduce the pressure from upward social comparison by showcasing relatable and genuine experiences.

### B. User Testing Procedure

The user testing session is designed to last 30-35 minutes per participant and includes three main phases: Pre-Experiment Setup, Controlled Task Phase, and Post-Experiment Feedback and Analysis.

**1. Pre-Experiment Setup (5 minutes)**:

- Participants (N = 50) will be recruited from the Korean MZ generation (aged 18-35), selected based on their active engagement with health-related content on social media.
- A **pre-experiment survey** will be conducted to measure baseline levels of social comparison tendencies, self-esteem, and typical social media usage patterns. Standardized scales, including the Social Comparison Scale and the Rosenberg Self-Esteem Scale, will be utilized.

**2. Controlled Task Phase (20-25 minutes)**:

- **Introduction and Training (5 minutes)**:

Participants will receive a brief orientation on the prototype, including an explanation of the "Inspiration" and "Reality" modes. They will be encouraged to use the Think-Aloud Method, verbalizing their thoughts, reactions, and decision-making processes throughout the session.

**Task Scenario 1: Content Exploration (5 minutes)**:
Participants will be asked to explore both modes freely and describe their impressions as they switch between "Inspiration" and "Reality." The Think-Aloud Method will capture their immediate reactions and thoughts about the differences between the modes.
Sample prompts: "What do you notice about the content in this mode?" and "How does this content make you feel?"

**Task Scenario 2: Emotional State Assessment (5 minutes)**:
Participants will be guided through scenarios designed to elicit specific emotional responses:
"Imagine you are feeling demotivated after a tough workout. Which mode would you choose to view, and why?"
"If you were feeling insecure about your fitness progress, which mode do you think would help you feel better?"
This task aims to assess the users' mode preferences based on their current emotional needs and to understand how the content affects their emotional state.

**Task Scenario 3: Decision-Making Task (5-10 minutes)**: Participants will be presented with a series of health-related posts and asked to decide which mode (Inspiration or Reality) they believe each post belongs to. This task tests their understanding of the mode distinctions and helps validate the effectiveness of the mode categorization.

During this task, participants will be asked to explain their reasoning out loud: "Why did you categorize this post as 'Inspiration' or 'Reality'?"

**3. Post-Experiment Feedback and Analysis (5-10 minutes)**:

**Post-Experiment Survey**: After the controlled tasks, participants will complete a survey to measure any changes in self-esteem, perceived authenticity of the content, and experiences of social comparison. This survey will include items from the Perceived Authenticity Scale, the Social Comparison Scale, and the User Engagement Scale.

**Brief Interview Session**: A semi-structured interview will follow the survey, allowing participants to provide additional feedback on their overall experience:

- "Which mode did you prefer and why?"
- "How did switching between modes affect your perception of the content?"
- "Do you have any suggestions for improving the toggle feature or the mode descriptions?"

## III. RESULTS

Based on the proposed dual-mode interface design and the literature review, we anticipate several key outcomes from the planned user testing sessions:

1. **Enhanced Perceived Authenticity**:

It is expected that the Reality Mode will be perceived as more authentic and relatable compared to the Inspiration Mode. Users may find the unfiltered, candid content in the Reality Mode more reflective of real-life health experiences, helping to reduce the pressure of unrealistic standards.

2. **Reduction in Negative Social Comparison**:

The toggle feature is anticipated to offer users greater control over their content consumption, allowing them to select a mode that aligns with their current emotional state. This flexibility may help mitigate the negative effects of upward social comparison, particularly when users opt to view content in the Reality Mode.

3. **Improved User Satisfaction**:

We expect that providing users with the choice between aspirational and relatable content will enhance overall user satisfaction. Users may appreciate the ability to tailor their feed based on their mood and preferences, which could lead to a more personalized and engaging experience.

4. **Varied Preferences Based on Emotional State**:

Users are likely to switch between modes depending on their emotional state and current needs. For example, users seeking motivation might prefer the Inspiration Mode, while those feeling self-conscious or in need of comfort may gravitate towards the Reality Mode. This behavior would support the hypothesis that the dual-mode interface caters to diverse user needs.

## IV. CONCLUSION

This pre-study outlines a novel dual-mode interface design aimed at mitigating the negative effects of social comparison on health-related social media content. By offering users the choice between **Inspiration Mode** and **Reality Mode**, the design seeks to balance motivation with authenticity, catering to the diverse needs of the Korean MZ generation.

The expected outcomes suggest that the Reality Mode will be perceived as more authentic and comforting, potentially reducing the pressures of upward comparison, while the Inspiration Mode may provide the motivational boost that some users seek. The ability to toggle between these modes is expected to enhance user satisfaction by offering a personalized and flexible content experience.

In conclusion, the proposed design represents a promising approach to addressing the challenges of social comparison in the digital health space. By empowering users to control their content experience, the dual-mode interface aligns with the principles of Healthy Pleasure, promoting a balanced and enjoyable pursuit of wellness. Future research should focus on validating these expected outcomes through user testing, exploring the integration of user-controlled content categorization, and evaluating the long-term impact of the design on user well-being.